\documentclass[12pt]{article}

\usepackage{epsfig}
\usepackage{amssymb,amsmath,amsfonts}
\usepackage{multirow}
\usepackage{rotating}
\usepackage{graphicx}
\usepackage{lscape}
\usepackage{hyperref}
\usepackage{ulem}
\usepackage{xcolor}

\begin{document}

\begin{flushright}
\end{flushright}

\def\g2{{g\hspace{-.2em}-\hspace{-.2em}2}}

\vspace*{1cm}
\begin{center}
\baselineskip 20pt 
{\Large\bf
Tau $\g2$ at $e^-e^+$ colliders \\
with momentum-dependent form factor
}

\vspace{1cm}

{\large
Hieu Minh Tran$^{a,}$\footnote{E-mail: hieu.tranminh@hust.edu.vn}
and 
Yoshimasa Kurihara$^{b,}$\footnote{E-mail: yoshimasa.kurihara@kek.jp} 

}

\vspace{0.5cm}

{\baselineskip 20pt \it
$^a$Hanoi University of Science and Technology, 1 Dai Co Viet Road, Hanoi, Vietnam\\
\vspace{2mm}
$^b$High Energy Accelerator Reseach Organization, Oho 1-1, Tsukuba, Ibaraki 305-0801, Japan\\
\vspace{2mm}
}

\end{center}

\vspace{1.5cm}
{\bf Abstract:}
The deviation between the prediction based on the standard model  and the measurement of the muon $\g2$ is currently at $3\hspace{-.2em}-\hspace{-.2em}4 \sigma$ (can be up to $7\sigma$ in the upcoming E989 experiment).
If this discrepancy is attributable to new physics, it is expected that the new contributions to the tau $\g2$ even larger than 
 those of muon due to its large mass. 
However, it is much more difficult to directly measure the tau $\g2$ because of its short lifetime.
In this report, we consider the effect of the tau $\g2$ at $e^-e^+$ colliders using a model independent approach.
Using the tau pair production channel at the Large Electron Position Collider (LEP), we have determined the allowed range for the new physics contribution of the tau $\g2$ assuming a $q^2$-dependence ansatz for the magnetic form factor.
In our analysis, we consider the standard model one-loop correction as well as the initial state photon radiation.
We also investigated the prospect at future $e^-e^+$ colliders, and determined the expected allowed range for the new physics contribution to the tau $\g2$.
Given the proposed beam polarization configuration at the International Linear Collider (ILC), we have analyzed the dependence of this allowed range on the integrated luminosity, as well as the relative systematic error.

\thispagestyle{empty}


\newpage

\baselineskip 18pt

\addtocounter{page}{-1}
\section{Introduction}


The discrepancy of approximately $3\hspace{-.2em}-\hspace{-.2em}4$ standard deviations 
\cite{Blum:2013xva,Keshavarzi:2018mgv,Davier_2020,Davier_2017,Davier_2011} 
between the prediction of the standard model (SM) and the experimental value of the muon anomalous magnetic moment, 
$a_\mu = {(g_\mu -2)}{/2}$, 
may be an indication of the limit of this theoretical model. 
This anomaly is being  investigated at the E989 experiment \cite{Grange:2015fou}.
If the measured center value of $a_\mu$ is still the same, the deviation will be confirmed at the  level of $7.0\sigma$ \cite{Keshavarzi:2018mgv}, which would strongly imply the involvement of new underlying physics coupled to leptons \cite{Tran:2018kxv}.
%
%
%
Assuming the universality of lepton, it is expected that the new physics contributions to  a lepton's anomalous magnetic moment are proportional to the squared ratio of its mass and the new physics scale, 
$\Delta a_l \sim {m_l^2}{/\Lambda_{NP}^2}$.
Therefore, the tau anomalous magnetic moment ($a_\tau$) is much more sensitive to this new physics compared to a muon due to its large mass.

The SM prediction of the tau $\g2$ is given by \cite{Eidelman:2007sb, Samuel:1990su}
\begin{eqnarray}
a_\tau^{\text{SM}} = (117721 \pm 5) \times 10^{-8}.
\end{eqnarray}
Any observation of significant deviation from this value is evidence of new physics beyond the SM.
Given  the short lifetime, based on current technology, taus cannot not be placed in a storage ring to measure their spin precession as in the case of muons.
Therefore, for practical reasons, $a_\tau$ must be measured by extracting information from collision data.
%
%
%
In this regard, several collaborations including OPAL \cite{Ackerstaff:1998mt}, L3 \cite{Acciarri:1998iv}, and DELPHI  \cite{Abdallah:2003xd}
investigated the allowed range for $a_\tau$  
using the data from the Large Electron Position Collider (LEP).
Among them, the most severe bounds were given by the DELPHI Collaboration using the process 
$e^-e^+ \rightarrow e^-e^+\tau^-\tau^+$. 
Based on this result, the limits for the non-standard contribution to $a_\tau$ were obtained by comparing the measured cross-section and the SM prediction:
\begin{eqnarray}
-0.052 < a_\tau^\text{NP} < 0.013 \qquad 
	(\text{95\% C.L.}) .
\label{delphi}
\end{eqnarray}
Using an effective Lagrangian method, the bounds on the contribution of new physics beyond the SM ($a_\tau^\text{NP}$) to the tau anomalous magnetic moment were extracted from the LEP and SLD data, mostly for the 
$e^-e^+ \rightarrow \tau^-\tau^+$ channel
\cite{GonzalezSprinberg:2000mk, Eidelman:2016aih}:
\begin{eqnarray}
 -0.007 < a_\tau^\text{NP}  < 0.004 \qquad  (2\sigma)  .
\label{eff_method}
\end{eqnarray}
These analyses are based on the idea that, at low energies, any new physics beyond the SM results in effective high-dimensional operators built with SM fields suppressed by a typical high energy scale $\Lambda_\text{NP}$. 
Similarly, limits extracted from other channels were also studied in Ref.
\cite{Escribano:1993pq}.


There have been several proposals to measure the tau anomalous magnetic moment and to utilize it as a probe for new physics at the LHC using various channels 
\cite{delAguila:1991rm, Samuel:1992fm, Atag:2010ja, dyndal2020anomalous, Hayreter:2013vna, Galon:2016ngp, Koksal:2017nmy, Fomin:2018ybj, Fu:2019utm, Beresford:2019gww}.
In the future, colliders such as the International Linear Collider (ILC) \cite{Djouadi:2007ik} will have more center-of-mass energies, higher luminosity, and the ability to control the polarization of the beam.
They will provide us with opportunities for precision tests to elucidate the viability of the SM as well as probing new physics.
The sensitivity required to measure $a_\tau$ in future colliders such as the Future Circular Collider (FCC), and the Compact Linear Collider (CLIC) has also been estimated in Refs. 
\cite{Tabares:2001xq, Billur:2013rva, Ozguven:2016rst, Chen:2018cxt, Koksal:2018env, Koksal:2018xyi, Gutierrez-Rodriguez:2019umw}.


On one hand, the momentum dependence of the form factors in the $\bar{\tau} \tau \gamma$ vertex is usually neglected for simplicity.
%
%
Namely, all the particles in the vertex (taus and photons) are assumed to be almost on-shell ($q_\tau^2 \approx m_\tau^2$ and 
$q_\gamma^2 \approx 0$).
In Ref. \cite{Bernabeu:2007rr}, the authors noted that for many high energy processes, the measured experimental parameter is the magnetic form factor rather than $a_\tau$ because the involvement of taus and/or photons is off-shell 
The report also revealed that the tau magnetic form factor can be measured at high luminosity B/Flavor factories. 
This leads to the consideration of the $q^2$-dependence effect of the form factor in the determination of experimental bounds for the anomalous magnetic dipole moment. 
This is especially important for processes in which the photon is highly off-shell (large $q^2$).
On the other hand, the SM contribution to the anomalous magnetic moment only emerges at the loop level, but not at the tree level.
Hence, the analyses at the next-to-leading order is necessary to reduce the relevant theoretical uncertainty.
This is particularly important when the new physics contribution is of the same order as that of the SM or smaller.

In this report, we investigate the tau anomalous magnetic moment using the channel $e^-e^+ \rightarrow \tau^-\tau^+$ while
the $q^2$-dependence of the magnetic moment form factor is considered, and
the cross-section calculation is performed at the next to leading order including both one-loop and initial state photon radiation (ISR) corrections.
We will show that the role of this $q^2$-dependence is very important in the determination of the bounds for $a_\tau^\text{NP}$.
We also consider the prospects for future $e^-e^+$ colliders in which the luminosity will be significantly improved compared to the LEP experiment, and the beam polarization will be feasible.
Our approach is model independent, and the results can be applied to the Compact Linear Collider (CLIC), the ILC, and the SuperKEKB experiments.

The structure of this report is as follows.
In Section 2, we briefly review the tau anomalous magnetic moment and the corresponding form factor.
In Section 3, using the $q^2$-dependent form factor, we extract the bounds for $a_\tau^\text{NP}$ using the LEP-II data in the 
$e^-e^+ \rightarrow \tau^-\tau^+$ channel.
In Section 4, the prospect for future $e^-e^+$ colliders is investigated considering the projected high luminosity and the initial beam polarization.
Finally, Section 5 is the conclusion.

\section{Tau anomalous magnetic moment}

The determination of the tau anomalous magnetic moment is based on its contribution to the $\bar{\tau} \tau \gamma$ vertex of  tau production processes or decays at colliders.
The general formula of such an interaction vertex between an off-shell photon with an arbitrary 4-momentum $q^\mu$ and two on-shell taus is given by \cite{Eidelman:2016aih}
\begin{eqnarray}
\Gamma^\mu (q^2) &=& 	- i e
	\left\lbrace
		\gamma^\mu F_1(q^2) + 
		\frac{\sigma^{\mu\nu} q_\nu}{2 m_\tau}
			\left[
			i F_2(q^2) + F_3(q^2) \gamma_5
			\right]	+
		\left(
			\gamma^\mu -
			\frac{2 q^\mu m_\tau}{q^2}
		\right)
		\gamma_5 F_4(q^2)
	\right\rbrace	.
\label{vertex}
\end{eqnarray}
where 
$\sigma^{\mu\nu} = 
	\frac{i}{2} [\gamma^\mu,\gamma^\nu]$.
The function $F_1(q^2)$ is the Dirac form factor that describes the electric charge distribution and satisfies the requirement $F_1(0) = 1$.
$F_2(q^2)$ and $F_3(q^2)$ are the form factors related to the magnetic and electric dipole moment respectively.
The last function $F_4(q^2)$ is the anapole form factor.
In the limit $q^2 = 0$, the form factors $F_2$ and $F_3$ result in the anomalous magnetic moment $a_\tau$ and the electric dipole moment $d_\tau$:
\begin{eqnarray}
F_2(0)	&=&	a_\tau	, \\
F_3(0)	&=&	- \frac{2 m_\tau d_\tau}{e} ,
\end{eqnarray}
The magnetic form factor $F_2(q^2)$ is dominated by the QED contribution that was computed at one loop in previous works 
\cite{Bernabeu:2007rr, Itzykson:1980rh}
\begin{eqnarray}
F_2^\text{QED}(q^2) &=&
	\left( \frac{\alpha}{2\pi} \right)
	\frac{2 m_\tau^2}{s}
	\frac{1}{\beta}
	\left(
	\text{log} \frac{1+\beta}{1-\beta} - i \pi
	\right)	,
\label{f2qed}
\end{eqnarray}
where $s = q^2 > 4 m_\tau^2$,
$\alpha$ is the fine structure constant, and
$\beta = \sqrt{1 - \frac{4 m_\tau^2}{s}}$ is the velocity of the $\tau$ lepton.
Hence, we have $a_\tau \approx a_\tau^\text{QED} = {\alpha}{/2\pi}$.

\begin{figure}[h]
\begin{center}
\includegraphics[scale=0.6]{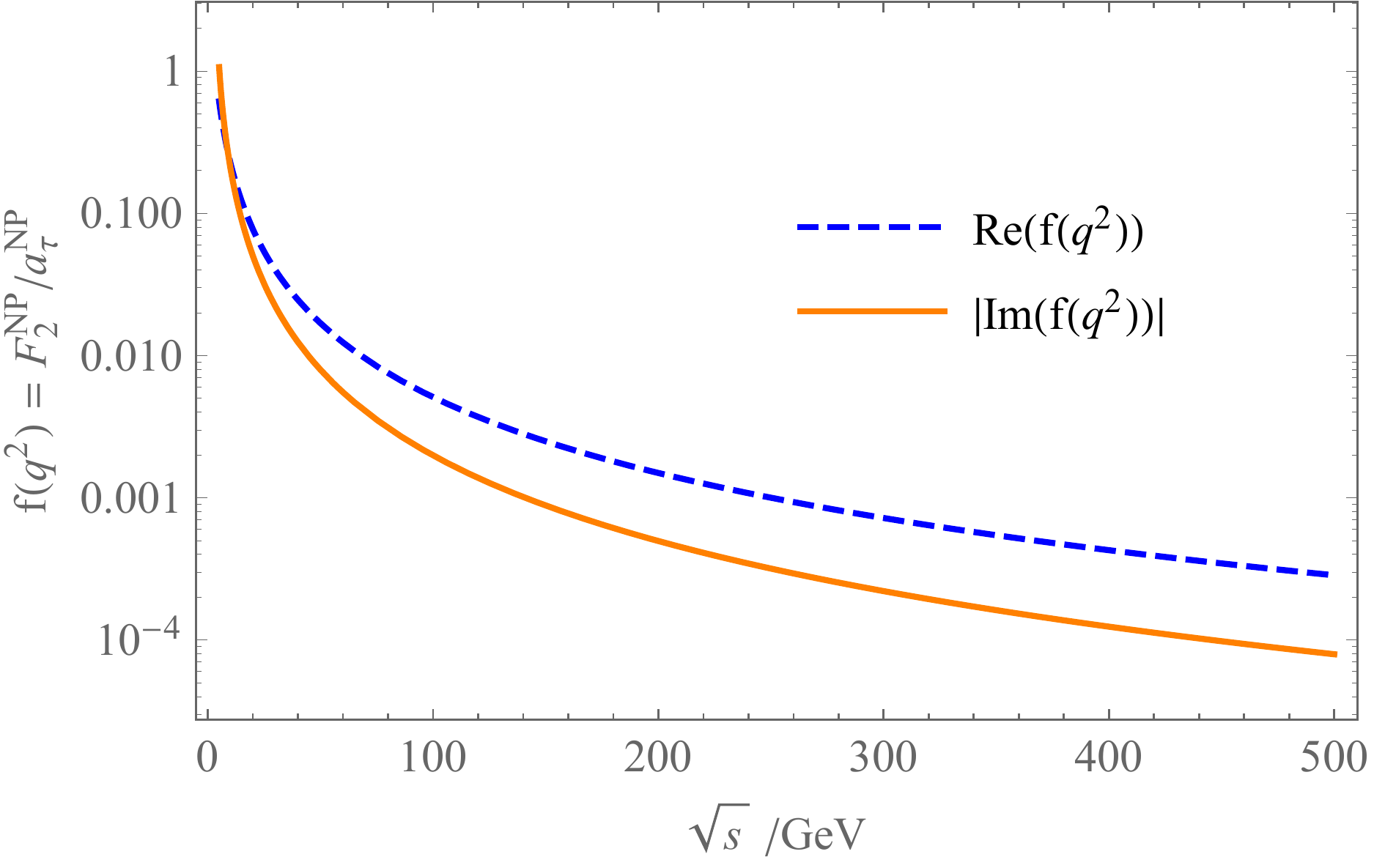}
  \caption{
The momentum dependence of the new physics contributions to the anomalous magnetic-moment form-factor according to the QED ansatz (\ref{f}). Here, we denote $s = q^2$.
}
\label{q-dependence}
\end{center}  
\end{figure}

In our analysis, we consider the SM one-loop calculation together with the initial state radiative (ISR) corrections.
Therefore, the SM contributions to the form factors in Eq. (\ref{vertex}) are considered.
For simplicity, we assume that new physics beyond the SM affects the cross-section of the tau-pair production process only
via its contributions to the form factor $F_2(q^2)$,
and that the $q^2$-dependence of such non-SM contributions follows the QED ansatz (\ref{f2qed}):
\begin{eqnarray}
F_2^\text{NP}(q^2) &=&
	a_\tau^\text{NP} f(q^2) ,
\label{f2NP}
\end{eqnarray}
where $a_\tau^\text{NP}$ is the new physics contribution to the tau anomalous magnetic moment, and the function $f(q^2)$ is given as
\begin{eqnarray}
f(q^2)	&=&
	\frac{2 m_\tau^2}{s}
	\frac{1}{\beta}
	\left(
	\text{log} \frac{1+\beta}{1-\beta} - i \pi
	\right)	.
\label{f}
\end{eqnarray}
Here, in the low momentum limit, we can see that
$F_2^\text{NP}(0) = a_\tau^\text{NP}$.
The behavior of the function $f(q^2)$ at larger $q^2$ is shown in Fig. \ref{q-dependence}.
Based on this parameterization, our approach to determine $a_\tau^\text{NP}$ is model independent.

\section{Bounds from LEP-II data}


In this section, we consider the process 
$e^-e^+ \rightarrow \tau^-\tau^+$ at the LEP experiment.
The experimental values and the corresponding errors of the cross-sections and the forward-backward asymmetries at various center-of-mass energies 
($\sqrt{s} =$ 130, 136, 161, 172, 183, 189, 192, 196, 200, 202, 205, 207 GeV)
are taken from Ref. 
\cite{Schael:2013ita}.
For simplicity, they are regarded as independent observable data to be fitted in our analysis.
The SM predictions for such quantities are calculated at the next-to-leading order, including the consideration of the one-loop corrections and the higher order resummation due to ISR effects \cite{Fujimoto:ISR}.
The non-SM contributions are implemented in the tree amplitudes by introducing the form factor 
$F_2^\text{NP}(q^2)$ in the 
$\bar{\tau}\tau\gamma$ vertex in Eq. (\ref{vertex}).
Therefore, $a_\tau^\text{NP}$ is the only free parameter in our theoretical analysis.
In the limit $a_\tau^\text{NP} \rightarrow 0$, the SM case is recovered.

\begin{figure}[h]
\begin{center}
\includegraphics[scale=0.6]{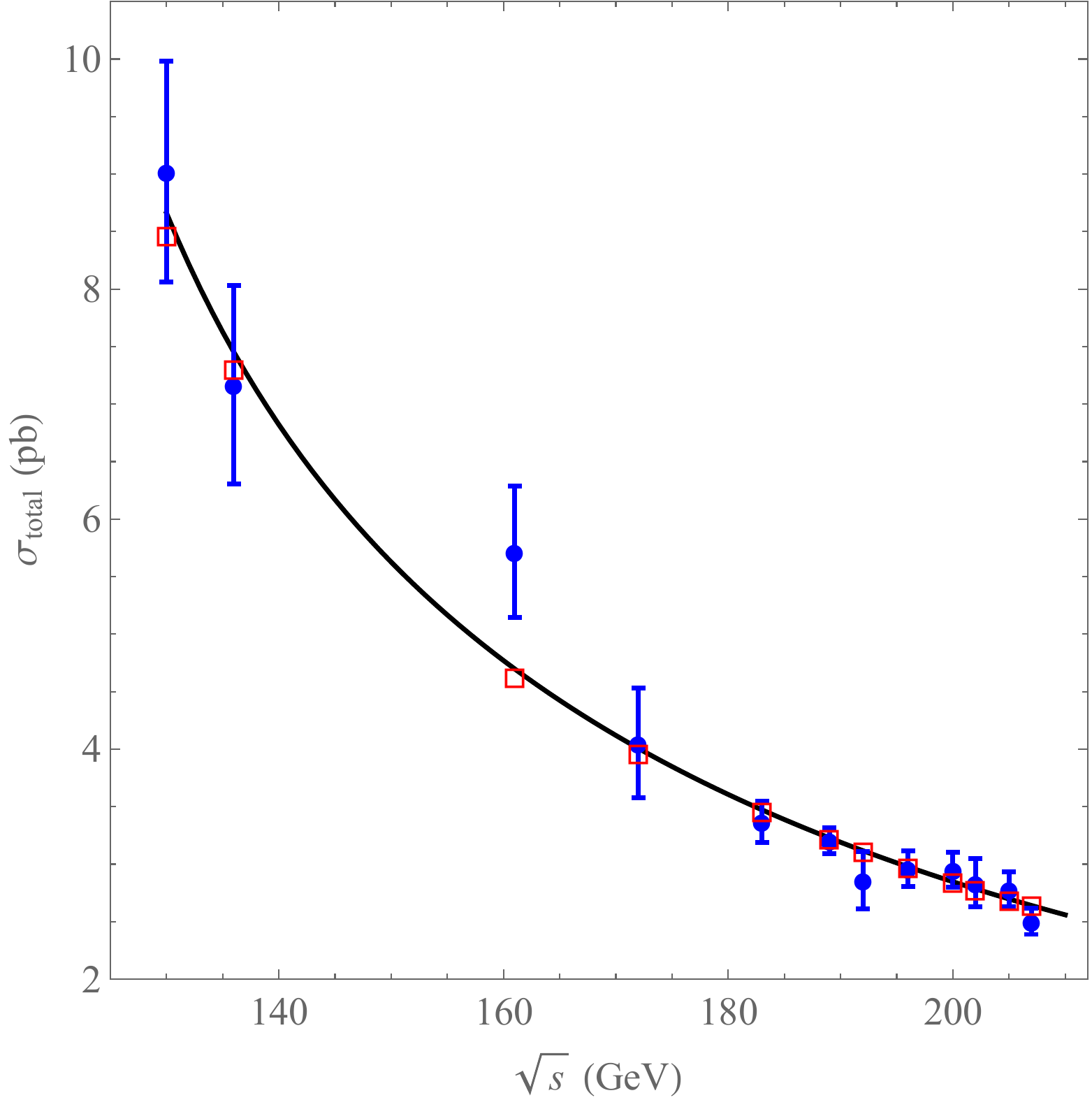}
  \caption{
The total cross-section of the process 
$e^-e^+ \rightarrow \tau^-\tau^+$ as a function of the center-of-mass energy $\sqrt{s}$.
The LEP-II data and the ZFITTER predictions for the cross-sections are shown  as dots with corresponding error bars and red empty squares, 
respectively \cite{Schael:2013ita} .
The SM calculation performed using GRACE at the next-to-leading order is represented by the black solid line where $a_\tau^\text{NP} = 0$.
}
\label{GZ}
\end{center}  
\end{figure}

For the considered numerical calculation, we employ the GRACE-Loop system that can evaluate the cross-sections and decay rates of physical processes for the SM \cite{Belanger:2003sd}
and its supersymmetric extensions \cite{Fujimoto:2007bn}.
This system is able to handle one-loop electroweak corrections for processes with two, three or four particles in the final state 
\cite{Belanger:2003ya}.
In this case, the renormalization of the electroweak interaction is performed using the on-shell scheme \cite{Aoki:1982ed},
and the infrared divergences are regulated by introducing a fictitious photon mass \cite{Fujimoto:1990tb}.
To investigate the Dirac and tensor algebra in $n$-dimensions, the symbolic manipulation package FORM \cite{Vermaseren:2000nd} is used.
The loop integrations are performed using the package FF \cite{vanOldenborgh:1990yc}
after reducing all tensor one-loop integrals to scalar integrals in a specific way \cite{Belanger:2003sd}.
The adaptive Monte Carlos integration package BASES \cite{Kawabata:1985yt}
is then used to perform the phase-space integrations.
The GRACE system uses the $R_\xi$ gauge for the linear gauge fixing terms, which should be checked using the additional non-linear gauge fixing Lagrangian 
\cite{Belanger:2003sd, Boudjema:1995cb} 
for consistency.
The final results are independent of fictitious parameters such as the photon mass and non-linear gauge parameters, which was numerically evaluated up to approximately fifteen digits against changing these values within several order of magnitudes at typical phase-space points.
The renormalization group running effect of the fine structure constant is 
considered in our calculation.
We compare the cross-section calculations for the $\tau$ pair production process in the SM with GRACE and ZFITTER \cite{Schael:2013ita} in Fig. \ref{GZ}.
In this case, the black solid line and the red empty squares correspond to the GRACE and ZFITTER theoretical calculations  respectively, whereas the LEP-II data are represented as blue points with error bars.
The figure shows that the GRACE and ZFITTER results are in a good agreement.

When $a_\tau^\text{NP}$ is switched on, the best-fit value of $a_\tau^\text{NP}$ is obtained by minimizing the $\chi^2$ function defined as
\begin{eqnarray}
\chi^2 (a_\tau^\text{NP})	
&=&
	\sum_{k=1}^{2n} \chi^2_k (a_\tau^\text{NP})		\nonumber	\\
&=&	
	\left[
	\sum_{i=1}^n 
		\left(	
		\frac{\sigma(a_\tau^\text{NP},s_i) - \sigma^\text{exp}(s_i)}{\Delta \sigma^\text{exp}(s_i)} 
		\right)^2	+		
	\sum_{i=1}^n 
		\left(
		\frac{A_\text{FB}(a_\tau^\text{NP},s_i) - A_\text{FB}^\text{exp}(s_i)}{\Delta A_\text{FB}^\text{exp}(s_i)}
		\right)^2
	\right]	,
\label{chi2}
\end{eqnarray}
where 
$\sigma^\text{exp}$, 
$A_\text{FB}^\text{exp}$, 
$\Delta \sigma^\text{exp}$,
and $\Delta A_\text{FB}^\text{exp}$ 
are the experimental values of cross-sections, forward-backward asymmetries and their corresponding errors respectively.
$\sigma$ and $A_\text{FB}$ denote the theoretical predictions for the cross-section and the forward-backward asymmetry, respectively.
$s_i$ represent the squared center-of-mass energies, and $n = 12$ is the number of collision energies examined in the LEP-II experiments
(see Table 3.4 in Ref. \cite{Schael:2013ita}).
Since the cross-section is a quadratic function of $a_\tau^\text{NP}$, the $\chi^2$ function is a quartic function of $a_\tau^\text{NP}$.
The likelihood function is determined from the $\chi^2$ function as
\begin{eqnarray}
\mathcal{L}(a_\tau^\text{NP})	&=&
	\frac{1}{N}
	e^{-\frac{1}{2}\chi^2(a_\tau^\text{NP})} \quad ,
\end{eqnarray}
where the normalization constant $N$ is defined as
\begin{eqnarray}
N	&=&
	\int\limits_{-\infty}^\infty 
	e^{-\frac{1}{2}\chi^2(a_\tau^\text{NP})}
	d a_\tau^\text{NP}	.
\end{eqnarray}
Equivalent to minimizing the $\chi^2$ function, the best fit value can also be obtained by maximizing the likelihood function.
In addition, a confident interval of $a_\tau^\text{NP}$ can be derived using the likelihood method as
\begin{eqnarray}
\int\limits_{-\infty}^{L} \mathcal{L}(a_\tau^\text{NP}) d a_\tau^\text{NP}	=	
\int\limits_{R}^{\infty} \mathcal{L}(a_\tau^\text{NP}) d a_\tau^\text{NP}	=	
\frac{1 - CL}{2}	,
\end{eqnarray}
where $CL$ is the confident level of the interval between the lower limit $L$ and the upper limit $R$ of $a_\tau^\text{NP}$.

\begin{figure}
\begin{center}
\includegraphics[scale=0.6]{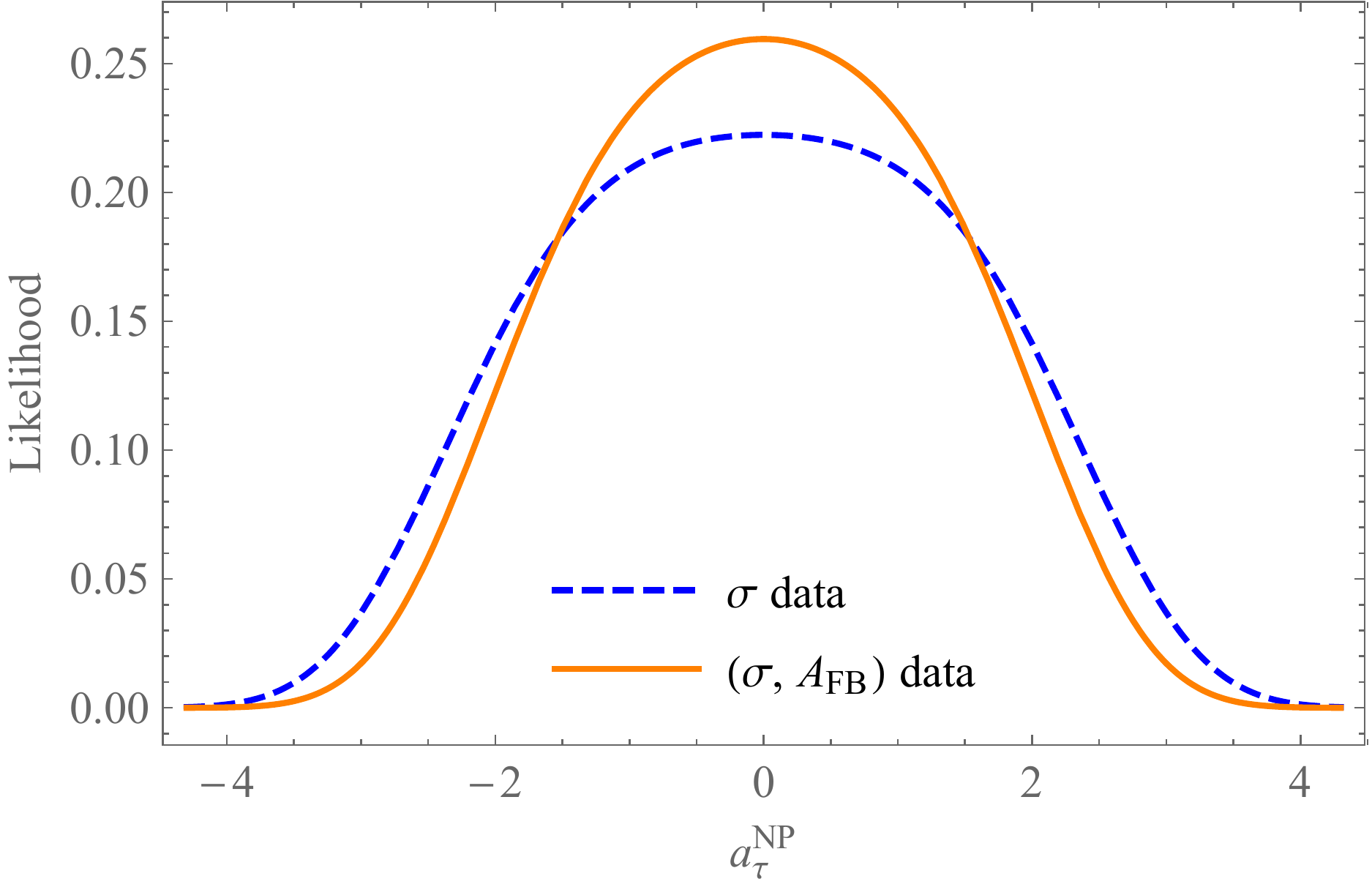}
  \caption{
The likelihood function of the new physics contribution to the tau anomalous magnetic moment, $\mathcal{L}(a_\tau^\text{NP})$. 
For the blue dashed line, the LEP-II data for only the cross-section are considered.
For the orange solid line, the LEP-II data for both the cross section and the forward-backward asymmetry are considered.
}
\label{LH}
\end{center}  
\end{figure}

In Fig. \ref{LH}, the likelihood functions are plotted for two cases: 
$(i)$ only LEP-II data on the total cross-section are considered (blue dashed line); and 
$(ii)$ LEP-II data on both the total cross section and the forward-backward asymmetry are considered (orange solid line).
We find the $95\%$ C.L. interval to be:
\begin{eqnarray}
-2.75 < a_\tau^\text{NP} < 2.75, 
\label{wide}
\end{eqnarray}
for the case $(i)$, and
\begin{eqnarray}
-2.46 < a_\tau^\text{NP} < 2.46,
\label{narrow}
\end{eqnarray}
for the case $(ii)$.
The best fit value for $a_\tau^\text{NP}$ is 0 for both cases, implying that the SM without exotic coupling is currently the best model using the LEP-II data of the $e^-e^+ \rightarrow \tau^-\tau^+$ process.
Comparing Eqs. (\ref{wide}) and (\ref{narrow}), we observe that the $95\%$ C.L. interval is narrowed by about 10\% when additional data on the forward-backward asymmetry are considered.

\begin{figure}[h]
\begin{center}
\includegraphics[scale=0.7]{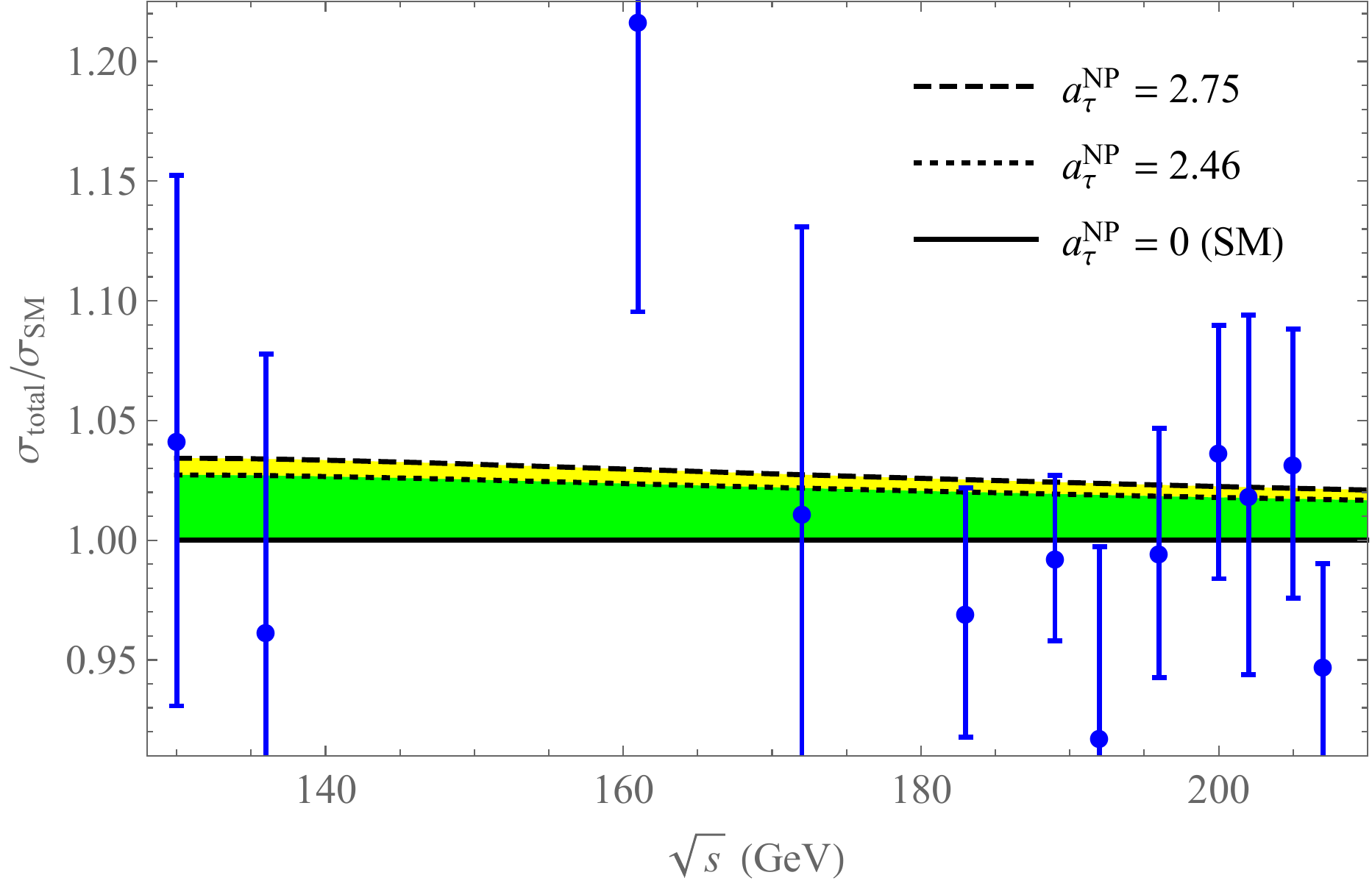}
  \caption{
The total cross-section normalized by the SM prediction of the process 
$e^-e^+ \rightarrow \tau^-\tau^+$ as a function of the center-of-mass energy $\sqrt{s}$.
The LEP-II data are shown in blue error bars
\cite{Schael:2013ita}.
The SM calculation performed using GRACE at the next-to-leading order is represented by the black solid line where $a_\tau^\text{NP} = 0$. 
The dashed and dotted lines correspond to the values $a_\tau^\text{NP} = 2.75$ and 2.46, respectively.
}
\label{crossX_n}
\end{center}  
\end{figure}

The total cross-section normalized by the SM cross-section is depicted in Fig. \ref{crossX_n} as a function of the center-of-mass energy.
In this case, the colored (both yellow and green) regions correspond to the interval 
in Eq. (\ref{wide}), 
whereas the green region corresponds to the interval in Eq. (\ref{narrow}).
The blue error bars are the LEP data on the normalized total cross-section.
Given that the cross-section strongly depends on the magnitude of $a_\tau^\text{NP}$ as a quartic function, the additional consideration of the data on the forward-backward asymmetry is important to reduce the width of the 95\% C.L. interval and the corresponding allowed region for the total cross-section.
In the future experiments with higher luminosity, we can expect that a more precise determination of the cross-section will result in a narrower allowed region for $a_\tau^\text{NP}$.

\section{Future prospect}

In future project such as the ILC, the center-of-mass energy will be fixed at 250 GeV for the first stage, and the corresponding integrated luminosity of $\mathcal{O}$(ab$^{-1}$) will significantly reduce the statistical uncertainties \cite{Barklow:2015tja} compared to the LEP experiment.
Moreover, the polarized initial $e^-$ and $e^+$ beams will enhance the capability for physics investigations such as testing the SM or identifying new particles and interactions.
The scattering cross-section with an arbitrary polarization combination is determined by 
\cite{MoortgatPick:2005cw}
\begin{eqnarray}
\sigma (P_{e^-}, P_{e^+}) &=&
	\frac{1}{4}
	\left[
	(1+P_{e^-})(1+P_{e^+})\sigma_\text{RR} +
	(1-P_{e^-})(1-P_{e^+})\sigma_\text{LL} 
	\right.		\nonumber \\
&&	\left.
+	(1+P_{e^-})(1-P_{e^+})\sigma_\text{RL} +
	(1-P_{e^-})(1+P_{e^+})\sigma_\text{LR} 
	\right] ,
\end{eqnarray}
where $P_{e^-}$ and $P_{e^+}$ are the polarization levels of the electron and positron beams with a range of values of $[-1,+1]$.
The cross-sections $\sigma_\text{RR,LL,RL,LR}$ correspond to the cases with 100\% beam polarization for both initial beams.
The current expectations for the magnitudes of the beam polarization levels are 80\% for electrons, and 30\% for positrons.  
In our analysis, we use the updated luminosity sharing between different beam polarization configurations for $e^-e^+$ collision as given in Table \ref{luminosity}.

\begin{table}
\begin{center}
\begin{tabular}{|c|cccc|}
\hline
Sign($P_{e^-},P_{e^+}$)	&	$(-,+)$	&	$(+,-)$	&	$(-,-)$	&	$(+,+)$	\\
\hline
Luminosity (fb$^{-1}$)	&	45\%	&	45\%	&	5\%	&	5\%	\\
\hline
\end{tabular}
\caption{Luminosity sharing between different polarization configurations for $\sqrt{s} = 250$ GeV.
The values are taken from Table 6 of Ref. \cite{Bambade:2019fyw}.
}
\label{luminosity}
\end{center}
\end{table}

As in the SM, the most important  contributions to the $e^- e^+ \rightarrow \tau^- \tau^+$ process are from the s-channels with virtual photon and $Z$-boson exchanges in our case.
Since the center-of-mass energy at the ILC is far from the $Z$-pole, the contribution of the photon exchange diagram is dominant over that of the Z-boson exchange diagram.
This results in only a small difference between the cross-sections $\sigma_\text{LR}$ and $\sigma_\text{RL}$.
Given that the SM gauge bosons only couple to particles of the same chirality at the tree level, the cross-sections $\sigma_\text{RR,LL}$ are negligible compared to $\sigma_\text{RL,LR}$.
In a similar way, the effective coupling related to the anomalous magnetic moment term in the $\bar{\tau}\tau\gamma$ vertex (Eq. (\ref{f2NP})) does not experience a difference among the polarizations of the incoming beams as long as they have opposite signs.
Therefore, while the left-right asymmetry ($A_\text{LR}$) does not contain information about the new effective coupling, the entire 90\% of the high luminosity of the ILC corresponding to 
$(P_{e^-}, P_{e^+}) = (\mp80\%,\pm30\%)$ (see Table \ref{luminosity})
will contribute to our analysis, resulting in a very small statistical uncertainty.

\begin{figure}[h]
\begin{center}
\includegraphics[scale=0.6]{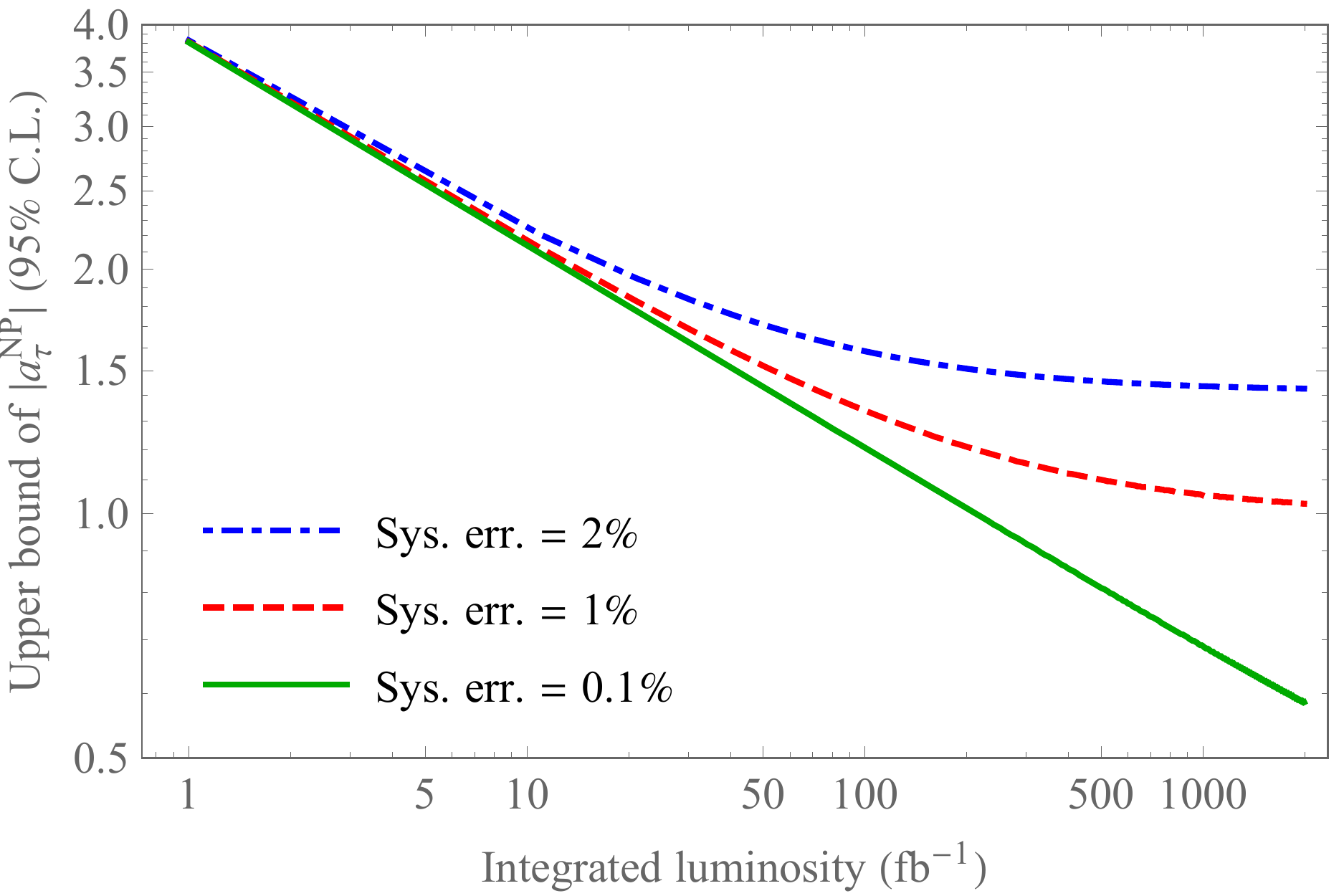}
  \caption{
The expected upper bound of $a_\tau^\text{NP}$ is plotted as a function of 
the integrated luminosity at the ILC for three cases in which the relative systematic errors are 2\%, 1\%, and 0.1\%, respectively.}
\label{upperbound_lum}
\end{center}  
\end{figure}

To investigate the prospect of $a_\tau^\text{NP}$ determination in the future ILC, we generate the measured angular distribution of the cross-section for each value of the integrated luminosity given a fixed relative systematic error.
The likelihood function and the 95\% C.L interval for $a_\tau^\text{NP}$ are then determined similarly to those in the previous section.
In Figure \ref{upperbound_lum}, we show the expected 95\% C.L. upper bound for 
$|a_\tau^\text{NP}|$ as a function of the integrated luminosity.
The blue (dash-dotted), red (dashed) and green (solid) lines correspond to relative systematic errors of 2\%, 1\% and 0.1\%, respectively.
We observe that the upper bound of $|a_\tau^\text{NP}|$ reduces quickly as the luminosity increases up to 100 fb$^{-1}$.
This is due to the suppression of the statistical uncertainty for larger luminosity.
For the integrated luminosity between 100 fb$^{-1}$ and 1000 fb$^{-1}$, the upper bounds that correspond to the relative systematic errors of 2\% and 1\% (the blue (dash-dotted) and red (dashed) lines, respectively) slowly decrease. 
They become saturated and are almost stable for luminosity values higher than 1000 fb$^{-1}$.
This is because the systematic uncertainty becomes dominant, and the reduction of the statistical uncertainty leads to a small reduction in the total uncertainty.
These relative systematic errors are of the same order as that for obtained the LEP experiment.
To effectively exploit the high luminosity for the ILC for shrinking of the allowed region of $a_\tau^\text{NP}$, the reduction of the systematic uncertainty is particularly important.
The case where the relative systematic error is approximately ten times smaller than that of the LEP experiment is considered.
In Figure \ref{upperbound_lum}, the green (solid) line corresponding to a relative systematic error of 0.1\% shows a further reduction of the upper bound of $|a_\tau^\text{NP}|$ when the luminosity is increased above 1000 fb$^{-1}$.
When the integrated luminosity reaches the ILC design-value of 2000 fb$^{-1}$, the allowed region for $a_\tau^\text{NP}$ is found to be
\begin{eqnarray}
-0.584 < a_\tau^\text{NP} < 0.584 \qquad (95\% \text{ C.L.}).
\label{ILCexpect1}
\end{eqnarray}
In this case, the systematic and the statistical uncertainties are of the same order.

\begin{figure}[h]
\begin{center}
\includegraphics[scale=0.6]{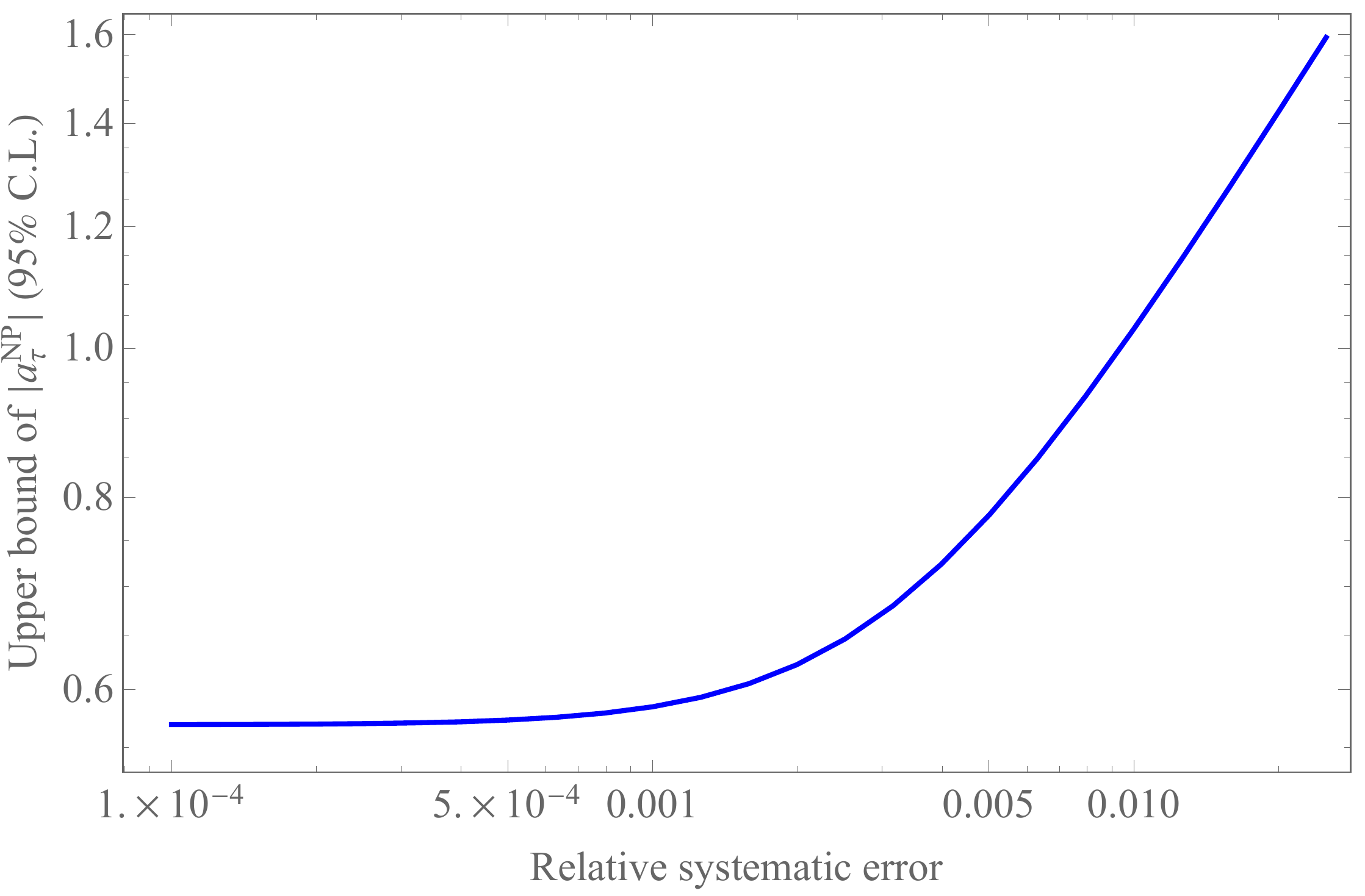}
  \caption{
The expected upper bound of $a_\tau^\text{NP}$ is plotted as a function of 
the systematic error for the case in which the integrated luminosity for the ILC is 2000 fb$^{-1}$.}
\label{sys_err}
\end{center}  
\end{figure}

The impact of the systematic uncertainty on the expected 95\% C.L. upper bound for 
$|a_\tau^\text{NP}|$ is depicted in Figure \ref{sys_err}, in which we fix the integrated luminosity to be 2000 fb$^{-1}$ according to the ILC design.
In this figure, we see that the upper bound can be reduced when the relative systematic error decreases from 2\% down to 0.1\%.
When the relative systematic error is below 0.1\%, the total uncertainty is dominated by the statistical uncertainty.
Therefore, a further reduction of the relative systematic error in this region does not lead to a significant decrease of the upper bound.
In fact, the upper bound becomes saturated for small values of the relative systematic error.
When the systematic error is as small as 0.01\%, 
for which the ratio between the systematic and statistical errors is of the similar order as that for the LEP experiment, the expected allowed region for the new physics contribution to the tau anomalous magnetic moment is found to be 
\begin{eqnarray}
-0.569 < a_\tau^\text{NP} < 0.569 \qquad (95\% \text{ C.L.}).
\label{ILCexpect2}
\end{eqnarray}
Comparing Eqs. (\ref{ILCexpect1}), (\ref{ILCexpect2}) and  (\ref{narrow}), we see that the expected allowed region obtained by the ILC experiment is approximately four times smaller than that obtained from the LEP experiment.

%
In this case, the theoretical calculation is assumed to be valid up to higher order corrections, and the theoretical uncertainty can be neglected for simplicity.
For such small experimental uncertainties as those at the ILC, in future projects, the higher loop corrections will be considered to reduce the relevant theoretical uncertainty such that it should be at least equivalent to the smallest experimental uncertainty (systematic or statistical ones), or much smaller in the best case.
For the relative systematic uncertainty of $0.1\%$
(which is comparable to the statistical uncertainty when the integrated luminosity is approximately 2000 fb$^{-1}$), $\mathcal{O}(\alpha^2)$
 corrections should be considered.
For a systematic uncertainty of $0.01\%$%
, both $\mathcal{O}(\alpha^2)$
 and 
$\mathcal{O}(\alpha^3)$
 corrections should be considered.

%
%

\section{Conclusion}

Although the anomalous magnetic moment of a muon can be determined precisely using the spin precession method with a muon storage ring, the measurement of this quantity is more challenging in the case of tau due to its short lifetime.
In the LEP experiment, tau $g-2$ was measured based on a simplified assumption that the $q^2$-dependence of the magnetic form factor is neglected.
Hence, it was argued that the results represent the bounds of the form factor rather than $a_\tau$.
In this report, assuming the QED ansatz for the $q^2$-dependent magnetic form factor, we have extracted the bounds for the new physics contribution to $a_\tau$ using the LEP-II data on the 
$e^-e^+ \rightarrow \tau^-\tau^+$ scattering process.
The $\mathcal{O}(\alpha)$ loop correction as well as the initial state photon radiation correction have been considered.
The 95\% C.L upper limit for 
$|a_\tau^\text{NP}|$ was determined to be 2.46.
We also investigated the prospect of $a_\tau^\text{NP}$ determination in future projects such as the ILC, in which a high luminosity and the possibility of initial beam polarization are the main advantages.
The dependences of the upper bound for $a_\tau^\text{NP}$ on the integrated luminosity and on the relative systematic error were analyzed.
Given the ILC updated design on the integrated luminosity and the beam polarization configuration, the expected 95\% C.L upper limit for 
$|a_\tau^\text{NP}|$ at the ILC is 0.569 , which is approximately four times better than the LEP bound.

\section*{Acknowledgement}

H.M.T. would like to thank the Institute of Particle and Nuclear Studies (IPNS) at KEK for their hospitality and support during his visit. 
The work of H.M.T is supported in part by the Vietnam National Foundation for Science and Technology Development (NAFOSTED) under grant number 103.01-2017.301.



\begin{thebibliography}{80}

\bibitem{Blum:2013xva} 
  T.~Blum, A.~Denig, I.~Logashenko, E.~de Rafael, B.~L.~Roberts, T.~Teubner and G.~Venanzoni,
  arXiv:1311.2198 [hep-ph].
  
\bibitem{Keshavarzi:2018mgv} 
  A.~Keshavarzi, D.~Nomura and T.~Teubner,
  Phys.\ Rev.\ D {\bf 97}, no. 11, 114025 (2018)
  [arXiv:1802.02995 [hep-ph]].

\bibitem{Davier_2020}
M.~Davier, A.~Hoecker, B.~Malaescu and Z.~Zhang,
Eur. Phys. J. C \textbf{80}, no.3, 241 (2020)
[arXiv:1908.00921 [hep-ph]].


\bibitem{Davier_2017}
M.~Davier, A.~Hoecker, B.~Malaescu and Z.~Zhang,
Eur. Phys. J. C \textbf{77}, no.12, 827 (2017)
[arXiv:1706.09436 [hep-ph]].



\bibitem{Davier_2011}
M.~Davier, A.~Hoecker, B.~Malaescu and Z.~Zhang,
Eur. Phys. J. C \textbf{71}, 1515 (2011)
[arXiv:1010.4180 [hep-ph]].



\bibitem{Grange:2015fou} 
  J.~Grange {\it et al.} [Muon g-2 Collaboration],
  arXiv:1501.06858 [physics.ins-det].

\bibitem{Tran:2018kxv} 
  See for example:
  H.~M.~Tran and H.~T.~Nguyen,
  Phys.\ Rev.\ D {\bf 99}, no. 3, 035040 (2019)
  [arXiv:1812.11757 [hep-ph]];
  N.~Okada and H.~M.~Tran,
  Phys.\ Rev.\ D {\bf 94}, no. 7, 075016 (2016)
  [arXiv:1606.05329 [hep-ph]],
  and references therein.
  
\bibitem{Eidelman:2007sb} 
  S.~Eidelman and M.~Passera,
  Mod.\ Phys.\ Lett.\ A {\bf 22}, 159 (2007)
  [hep-ph/0701260].

\bibitem{Samuel:1990su} 
  M.~A.~Samuel, G.~w.~Li and R.~Mendel,
  Phys.\ Rev.\ Lett.\  {\bf 67}, 668 (1991)
  Erratum: [Phys.\ Rev.\ Lett.\  {\bf 69}, 995 (1992)].

\bibitem{Ackerstaff:1998mt} 
  K.~Ackerstaff {\it et al.} [OPAL Collaboration],
  Phys.\ Lett.\ B {\bf 431}, 188 (1998)
  [hep-ex/9803020].

\bibitem{Acciarri:1998iv} 
  M.~Acciarri {\it et al.} [L3 Collaboration],
  Phys.\ Lett.\ B {\bf 434}, 169 (1998).

\bibitem{Abdallah:2003xd} 
  J.~Abdallah {\it et al.} [DELPHI Collaboration],
  Eur.\ Phys.\ J.\ C {\bf 35}, 159 (2004)
  [hep-ex/0406010].

\bibitem{GonzalezSprinberg:2000mk} 
  G.~A.~Gonzalez-Sprinberg, A.~Santamaria and J.~Vidal,
  Nucl.\ Phys.\ B {\bf 582}, 3 (2000)
  [hep-ph/0002203].

\bibitem{Eidelman:2016aih} 
  S.~Eidelman, D.~Epifanov, M.~Fael, L.~Mercolli and M.~Passera,
  JHEP {\bf 1603}, 140 (2016)
  [arXiv:1601.07987 [hep-ph]].




\bibitem{Escribano:1993pq} 
  R.~Escribano and E.~Masso,
  Phys.\ Lett.\ B {\bf 301}, 419 (1993);
  R.~Escribano and E.~Masso,
  Nucl.\ Phys.\ B {\bf 429}, 19 (1994)
  [hep-ph/9403304].
  
  
  

\bibitem{delAguila:1991rm} 
  F.~del Aguila, F.~Cornet and J.~I.~Illana,
  Phys.\ Lett.\ B {\bf 271}, 256 (1991).

\bibitem{Samuel:1992fm} 
  M.~A.~Samuel and G.~Li,
  Int.\ J.\ Theor.\ Phys.\  {\bf 33}, 1471 (1994).

\bibitem{Atag:2010ja} 
  S.~Atag and A.~A.~Billur,
  JHEP {\bf 1011}, 060 (2010)
  [arXiv:1005.2841 [hep-ph]].

\bibitem{dyndal2020anomalous}
M.~Dyndal, M.~Klusek-Gawenda, M.~Schott and A.~Szczurek,
[arXiv:2002.05503 [hep-ph]].


\bibitem{Hayreter:2013vna} 
  A.~Hayreter and G.~Valencia,
  Phys.\ Rev.\ D {\bf 88}, no. 1, 013015 (2013)
  Erratum: [Phys.\ Rev.\ D {\bf 91}, no. 9, 099902 (2015)]
  [arXiv:1305.6833 [hep-ph]].

\bibitem{Galon:2016ngp} 
  I.~Galon, A.~Rajaraman and T.~M.~P.~Tait,
  JHEP {\bf 1612}, 111 (2016)
  [arXiv:1610.01601 [hep-ph]].

\bibitem{Koksal:2017nmy} 
  M.~K\"{o}ksal, S.~C.~\'{I}nan, A.~A.~Billur, Y.~\"{O}zgüven and M.~K.~Bahar,
  Phys.\ Lett.\ B {\bf 783}, 375 (2018)
  [arXiv:1711.02405 [hep-ph]].

\bibitem{Fomin:2018ybj} 
  A.~S.~Fomin, A.~Y.~Korchin, A.~Stocchi, S.~Barsuk and P.~Robbe,
  JHEP {\bf 1903}, 156 (2019)
  [arXiv:1810.06699 [hep-ph]].

\bibitem{Fu:2019utm} 
  J.~Fu, M.~A.~Giorgi, L.~Henry, D.~Marangotto, F.~M.~Vidal, A.~Merli, N.~Neri and J.~Ruiz Vidal,
  Phys.\ Rev.\ Lett.\  {\bf 123}, no. 1, 011801 (2019)
  [arXiv:1901.04003 [hep-ex]].

\bibitem{Beresford:2019gww} 
  L.~Beresford and J.~Liu,
  arXiv:1908.05180 [hep-ph].


\bibitem{Djouadi:2007ik} 
  A.~Djouadi {\it et al.} [ILC Collaboration],
  arXiv:0709.1893 [hep-ph];
  H.~Baer {\it et al.},
  arXiv:1306.6352 [hep-ph];
  H.~Aihara {\it et al.} [ILC Collaboration],
  arXiv:1901.09829 [hep-ex];
  P.~Bambade {\it et al.},
  arXiv:1903.01629 [hep-ex];
  K.~Fujii {\it et al.} [LCC Physics Working Group],
  arXiv:1908.11299 [hep-ex].


\bibitem{Tabares:2001xq} 
  L.~Tabares and O.~A.~Sampayo,
  Phys.\ Rev.\ D {\bf 65}, 053012 (2002)
  [hep-ph/0111081].

\bibitem{Billur:2013rva} 
  A.~A.~Billur and M.~Koksal,
  Phys.\ Rev.\ D {\bf 89}, no. 3, 037301 (2014)
  [arXiv:1306.5620 [hep-ph]].

\bibitem{Ozguven:2016rst} 
  Y.~\"{O}zg\"{u}ven, S.~C.~\'{I}nan, A.~A.~Billur, M.~K\"{o}ksal and M.~K.~Bahar,
  Nucl.\ Phys.\ B {\bf 923}, 475 (2017)
  [arXiv:1609.08348 [hep-ph]].

\bibitem{Chen:2018cxt} 
  X.~Chen and Y.~Wu,
  arXiv:1803.00501 [hep-ph].

\bibitem{Koksal:2018env} 
  M.~K\"{o}ksal, A.~A.~Billur, A.~Guti\'{e}rrez-Rodríguez and M.~A.~Hern\'{a}ndez-Ru\'{i}z,
  Phys.\ Rev.\ D {\bf 98}, no. 1, 015017 (2018)
  [arXiv:1804.02373 [hep-ph]].

\bibitem{Koksal:2018xyi} 
  M.~K\"{o}ksal,
  J.\ Phys.\ G {\bf 46}, 065003 (2019)
  [arXiv:1809.01963 [hep-ph]].

\bibitem{Gutierrez-Rodriguez:2019umw} 
  A.~Gutiérrez-Rodríguez, M.~K\"{o}ksal, A.~A.~Billur and M.~A.~Hernández-Ruíz,
  arXiv:1903.04135 [hep-ph].








\bibitem{Bernabeu:2007rr} 
  J.~Bernabeu, G.~A.~Gonzalez-Sprinberg, J.~Papavassiliou and J.~Vidal,
  Nucl.\ Phys.\ B {\bf 790}, 160 (2008)
  [arXiv:0707.2496 [hep-ph]].

\bibitem{Itzykson:1980rh} 
  C.~Itzykson and J.~B.~Zuber,
  "Quantum Field Theory",
  New York, Usa: Mcgraw-hill (1980) 705 P.(International Series In Pure and Applied Physics)


\bibitem{Schael:2013ita} 
  S.~Schael {\it et al.} [ALEPH and DELPHI and L3 and OPAL and LEP Electroweak Collaborations],
  Phys.\ Rept.\  {\bf 532}, 119 (2013)
  [arXiv:1302.3415 [hep-ex]].

\bibitem{Fujimoto:ISR}
J.~Fujimoto, Y.~Kurihara and N.M.U.~Quach,  
Eur.\ Phys.\ J.\ C {\bf 79}, 506 (2019).

\bibitem{Belanger:2003sd} 
  G.~B\'{e}langer, F.~Boudjema, J.~Fujimoto, T.~Ishikawa, T.~Kaneko, K.~Kato and Y.~Shimizu,
  Phys.\ Rept.\  {\bf 430}, 117 (2006)
  [hep-ph/0308080].
  
\bibitem{Fujimoto:2007bn} 
  J.~Fujimoto, T.~Ishikawa, Y.~Kurihara, M.~Jimbo, T.~Kon and M.~Kuroda,
  Phys.\ Rev.\ D {\bf 75}, 113002 (2007);
  H.~M.~Tran, T.~Kon and Y.~Kurihara,
  Mod.\ Phys.\ Lett.\ A {\bf 26}, 949 (2011)
  [arXiv:1012.1730 [hep-ph]];
  Y.~Kouda, T.~Kon, Y.~Kurihara, T.~Ishikawa, M.~Jimbo, K.~Kato and M.~Kuroda,
  PTEP {\bf 2018}, no. 8, 083B03 (2018)
  [arXiv:1807.01911 [hep-ph]].

\bibitem{Belanger:2003ya} 
  G.~B\'{e}langer, F.~Boudjema, J.~Fujimoto, T.~Ishikawa, T.~Kaneko, Y.~Kurihara, K.~Kato and Y.~Shimizu,
  Phys.\ Lett.\ B {\bf 576}, 152 (2003)
  [hep-ph/0309010];
  G.~B\'{e}langer, F.~Boudjema, J.~Fujimoto, T.~Ishikawa, T.~Kaneko, K.~Kato, Y.~Shimizu and Y.~Yasui,
  Phys.\ Lett.\ B {\bf 571}, 163 (2003)
  [hep-ph/0307029];
  G.~B\'{e}langer, F.~Boudjema, J.~Fujimoto, T.~Ishikawa, T.~Kaneko, K.~Kato and Y.~Shimizu,
  Nucl.\ Phys.\ Proc.\ Suppl.\  {\bf 116}, 353 (2003)
  [hep-ph/0211268];
  K.~Kato, F.~Boudjema, J.~Fujimoto, T.~Ishikawa, T.~Kaneko, Y.~Kurihara, Y.~Shimizu and Y.~Yasui,
  PoS HEP {\bf 2005}, 312 (2006).
  
\bibitem{Aoki:1982ed} 
  K.~I.~Aoki, Z.~Hioki, M.~Konuma, R.~Kawabe and T.~Muta,
  Prog.\ Theor.\ Phys.\ Suppl.\  {\bf 73}, 1 (1982).

\bibitem{Fujimoto:1990tb} 
  J.~Fujimoto, M.~Igarashi, N.~Nakazawa, Y.~Shimizu and K.~Tobimatsu,
  Prog.\ Theor.\ Phys.\ Suppl.\  {\bf 100}, 1 (1990).

\bibitem{Vermaseren:2000nd} 
  J.~A.~M.~Vermaseren,
  math-ph/0010025.

\bibitem{vanOldenborgh:1990yc} 
  G.~J.~van Oldenborgh,
  Comput.\ Phys.\ Commun.\  {\bf 66}, 1 (1991).

\bibitem{Hahn:1998yk} 
  T.~Hahn and M.~Perez-Victoria,
  Comput.\ Phys.\ Commun.\  {\bf 118}, 153 (1999)
  [hep-ph/9807565].

\bibitem{Kawabata:1985yt} 
  S.~Kawabata,
  Comput.\ Phys.\ Commun.\  {\bf 41}, 127 (1986);
  S.~Kawabata,
  Comput.\ Phys.\ Commun.\  {\bf 88}, 309 (1995).

\bibitem{Boudjema:1995cb} 
  F.~Boudjema and E.~Chopin,
  Z.\ Phys.\ C {\bf 73}, 85 (1996)
  [hep-ph/9507396].


\bibitem{Barklow:2015tja} 
  T.~Barklow, J.~Brau, K.~Fujii, J.~Gao, J.~List, N.~Walker and K.~Yokoya,
  arXiv:1506.07830 [hep-ex].




\bibitem{MoortgatPick:2005cw}
G.~Moortgat-Pick {\it et al.}, 
Phys.\ Rept.\  \textbf{460}, 131-243 (2008)
[arXiv:hep-ph/0507011 [hep-ph]].


\bibitem{Bambade:2019fyw} 
  P.~Bambade {\it et al.},
  arXiv:1903.01629 [hep-ex];
  K.~Fujii {\it et al.} [LCC Physics Working Group],
  arXiv:1908.11299 [hep-ex].


  
  



%
%
  

\end{thebibliography}
\end{document}